\documentclass[%
 reprint,
nofootinbib,
 amsmath,amssymb,
 aps,
]{revtex4-1}

\usepackage{graphicx}
\usepackage{dcolumn}
\usepackage{bm}

\usepackage{physics}
\usepackage{xcolor} 

\usepackage[normalem]{ulem}
\usepackage{placeins}

\newcommand{\aap}{Astron. Astrophys.}

\newcommand{\jcap}{J. Cosmol. Astropart. Phys.}

\usepackage{graphicx}
\usepackage{amsmath,amsbsy,pifont, amssymb}
\usepackage{multirow}
\usepackage{soul, xcolor}

\usepackage{hyperref}
\newcommand{\comment}[1]{{}}

\newcommand{\lcdm}{\ensuremath{\Lambda\mathrm{CDM}}} 
\newcommand{\LCDM}{\ensuremath{\Lambda\mathrm{CDM}}}

 \renewcommand{\vec}[1]{\mathbf{#1}}

 \def\be{\begin{equation}}
\def\ee{\end{equation}}
\def\ba{\begin{eqnarray}}
\def\ea{\end{eqnarray}}

\newcommand{\class}{\textsc{class}}
 
\begin{document}

\preprint{APS/123-QED}

\title{Exploration of the Pre-recombination Universe with a High-Dimensional Model of an Additional Dark Fluid}

\author{Michael Meiers}
\email{mcmeiers@ucdavis.edu}
\author{Lloyd Knox}%
 \email{lknox@ucdavis.edu}
\affiliation{%
Department of Physics and Astronomy, University of California Davis, One Shields Avenue, Davis, California 95616
}%
\author{Nils Sch\"oneberg}
 \email{nils.science@gmail.com}
\affiliation{Institut de Ciències del Cosmos (ICCUB), Facultat de F\'isica, Universitat de Barcelona (IEEC-UB), Mart\'i i Franqués, 1, E08028 Barcelona, Spain}

\date{\today}

\begin{abstract}

We implement and explore high-dimensional generalized dark matter (HDGDM) with an arbitrary equation of state as a function of redshift as an extension to \lcdm. Exposing this model to CMB, BAO, and supernova data, we demonstrate that the use of marginalized posterior distributions can easily lead to misleading conclusions on the viability of a high-dimensional model such as this one. We discuss such pitfalls and corresponding mitigation strategies, which can be used to search for an observationally favored region of the parameter space. We further show that the HDGDM model in particular does show promise in terms of its ability to ease the Hubble tension once such techniques are employed, and we find interesting features in the best-fitting equation of state that can serve as an inspiration for future model building.

\end{abstract}

\maketitle

\section{Introduction}
Intensity and polarization maps of the cosmic microwave background (CMB) are highly sensitive to forces affecting the plasma in the two decades of scale factor evolution prior to recombination \cite[e.g.][]{Hu:1996mn,knox20}. As such, measurements of these maps are capable of constraining fairly high-dimensional models of an additional dark component that is dynamically important during that epoch. To explore the constraints from data on this epoch we extend the \lcdm\ model to include a fluid component modeled as ``generalized dark matter" \cite{Hu:1998kj} with viscosity and sound-speed parameters as free (time-independent) parameters and an equation of state parameter $w_{\rm g} \equiv P_{\rm g}/\rho_g$ that is a (approximately) free function of redshift. We refer to this model as high-dimensional generalized dark matter, or HDGDM.

This is a purely phenomenological model intended to enable a search for a variety of potential departures from \lcdm\ to see, given cosmological data, what departures are preferred, what are allowed, and what are ruled out. A discovery of a preference for a departure from \lcdm\ might then serve as guidance for future, more physical, model building. 

Our search procedure is a ``bottom-up" approach, driven by the data, distinguished from the analyses of lower-dimensional and more theoretically motivated models in \cite{Poulin:2018cxd,Agrawal:2019lmo, Smith:2020rxx,Hill:2020osr,Murgia:2020ryi,Niedermann:2020dwg,Aloni:2021eaq,Poulin:2023lkg,Hart:2022agu,Schoneberg:2022grr} which might instead be called ``top-down." In addition to presenting our HDGDM model, we present some initial lessons about use of such high-dimensional models for exploration of the implications of data. 

The HDGDM model is similar to a model used in two prior reconstructions of, in one case the expansion rate, and in another the dark energy fraction, as functions of redshift. In \cite{Hojjati:2013oya} the authors used a GDM model with its density history determined by its parameterized values at a set of control points in redshift to reconstruct $H(z)$ from the first release of cosmological data from {\it Planck} \cite{Planck:2013win}. More recently, the authors of \cite{Moss:2021obd} used a very similar GDM model to reconstruct the fraction of energy density in dark energy from $z=0$ to $z=10^5$ using various combinations of cosmological data sets. 

We find that exploration of high-dimensional models brings to the fore some issues of data analysis and inference that are not as present in the more commonly-explored lower-dimensional models. First and foremost, there are challenges that arise from the ``curse of dimensionality." Standard techniques for determining joint and marginal posterior probability distributions can end up taking a prohibitively long time to converge. 

Even if one can perform Bayesian inference in a high-dimensional model space, an additional pitfall remains that perfectly good regions of parameter space can effectively get lost, overwhelmed by much greater volume in other regions of parameter space. Such volume effects have been noted in lower-dimensional models too, and in models of early dark energy (EDE) in particular (see the recent EDE review \cite{Poulin:2023lkg} and references therein). The effective obscuration of regions that provide perfectly good fits leads us to a conclusion that if our goal is to find new and interesting ways of fitting the data, marginalization over vast regions of parameter space is probably not what we want to be doing.

We ground this discussion of the perils of exploring high-dimensional model spaces with an application of our HDGDM model inspired by the ``Step" model of \cite{Aloni:2021eaq}. In this model a light relic becomes non-relativistic and then decays into a massless particle. The equation-of-state parameter for this component thus begins as $w=1/3$, drops a bit, and then returns to $w=1/3$. They find some preference for such a component with the transition near $z \simeq 20\,000$. Interestingly, they find that this region of parameter space also leads to a higher $H_0$ and thus a reduction of the tension with the most recent measurement by the SH$_0$ES Collaboration \cite{Riess:2021jrx}. 

We use the HDGDM model to further explore this hint that a non-trivial evolution near \mbox{$z \sim 20\,000$} can have a strong impact on the Hubble tension. For this, we use a component that at both higher $z$ and lower $z$ is evolving like a light relic. Depending on how broadly we open up the prior constraints on $w_g$\,, the interesting solution of \cite{Aloni:2021eaq} can effectively disappear from view. 

We also find in this analysis the curious result that {\em adding} information (by restricting the allowed range of $w_g$ values), leads to a {\em broader} marginal posterior probability distribution for $H_0$\,. We attribute this unusual behavior to volume effects arising from implicit priors. Interpretation of results thus requires grappling with this dependence on priors. We also discuss a few mitigation strategies that can allow for effective model and parameter searching even in this high-dimensional parameter space. In particular, we outline the importance of likelihood minimization and the corresponding challenges.

In Section II we introduce the broad details of our HDGDM model and its implementation as an extension to the publicly available Einstein-Boltmann solver, \class{}. We also present some validation of this numerical code via comparison with existing \class{} capabilities. In Section III we present our exploration of the step-like parameter space, and lessons learned. We conclude in section IV.

\section{High-dimensional Generalized Dark Matter}

In the following subsections, we introduce critical features of a GDM model and present a realization of the model for numeric computation in \class{} as well our validation of this implementation.

\subsection{A Brief Introduction to Generalized Dark Mater}

Our high-dimensional modeling uses generalized dark matter (GDM), conceived initially by W. Hu \cite{Hu:1996}. In this subsection, we summarize the essential details of this model. For a deeper dive into the topic and to explore related theories, see \cite{Kopp}. 

\subsubsection{General Tensor Field Description}

We take the general decomposition of an energy-momentum tensor as 
\begin{equation}
    {T^\mu}_\nu= \rho\qty(u^\mu u_\nu)+\qty(u^\mu q_nu +q^\mu u_\nu) + P(u^\mu u_\nu+\delta^\mu_\nu)+ {\Sigma^\mu}_\nu~,
\end{equation} where $\rho$, $P$, $u^\mu$, $q^\mu$ and $\Sigma^\mu_\nu$ are the energy density, pressure, four-velocity, heat flux, and anisotropic stress respectively. The latter two are orthogonal to the four-velocity and vanish in an unperturbed FRWL spacetime. The anisotropic stress is traceless and can be non-zero at perturbative levels. Our work uses the Landau-Lifshitz frame, which sets the four-velocity so that the heat flux vanishes at all perturbative levels. 

The equation-of-state parameter $w \equiv P/\rho$ sets the background relationship between the pressure and the energy density. Traditional treatments restrict to a constant $w$, but it is a time-dependent function in general cases such as GDM. Perturbations have an adiabatic speed of sound set by 
\begin{equation}
    c_a^2\equiv\frac{\dot{P}}{\dot{\rho}}=w-\frac{\dot{w}}{3\mathcal{H}\qty(1+w)}~,
\end{equation}
where the dot denotes the derivative with respect to conformal time and $\mathcal{H} \equiv \dot{a}/a$. The equality is purely a consequence of energy-momentum conservation $\nabla_\mu{T^\mu}_\nu=0$, which requires that 
\begin{equation}
    \dot{\rho}=-3\mathcal{H}(1+w)\rho
\end{equation}
is separately satisfied for each non-interacting component of the universe. Note that $c_a^2\neq w$ unless we have constant~$w$. When subscripts are left out, we generally refer to the GDM species. However, we will include subscripts when ambiguity may arise.

\subsubsection{GDM Closure Relations}

At a perturbative level, two modifications to the standard treatment of perfect fluids close the evolution equations of GDM. The first is the inclusion of non-adiabatic pressure (NAP) contributions denoted $\Pi_\mathrm{nap} = \Pi - c_a^2\delta$ where $\Pi$ is the isotropic pressure perturbation defined by the trace of the spacelike component of the energy-momentum tensor ${T^i}_i=3\bar{\rho}(w+\Pi)$ and $\delta = \delta\rho /\bar{\rho}$ with $\bar{\rho}$ the background energy density. 

The equation of state parameter alone is insufficient to specify the NAP contributions, and models must specify this term as an additional input. As the NAP contributes in a gauge invariant way, any choices must respect this property. Hu's original proposal has $\Pi_{\rm nap}\propto\delta$ in the rest frame of the material, which is lifted to a gauge-invariant statement in Fourier space as \cite{Kopp};
\begin{align}
    \Pi_\mathrm{nap}=(c_s^2-c_a^2)\hat\Delta~, \\
    \hat\Delta= \delta +3(1+w)\mathcal{H}\theta/k^2~,
\end{align}
where the free parameter, $c_s^2$, can be thought of as the rest frame speed of sound. The divergence of velocity perturbation $\theta$ is defined to satisfy $\bar{\rho}\qty(1+w)\theta =ik^i\delta T^0_i$. In the perturbation's rest frame $\theta=0$, we see that $\hat{\Delta}=\delta$. Thus, the general expression $\Pi=c_a^2\delta+\Pi_\mathrm{nap}=c_a^2\delta+(c_s^2-c_a^2)\hat\Delta$ reduces to $\Pi=c_s^2\delta_{\rm rest}$ which supports the identification of $c_s^2$ as an effective speed of sound in the rest frame.

The second modeling choice supports growing modes for anisotropic stress. Here we differ slightly from Hu's original suggestion and use a change proposed by \cite{Kopp}. In  \cite{Ma:1995ey} the fourier space  scalar anisotropic stress perturbation, $\sigma$, is defined by  
\begin{equation}
    \sigma = -\frac{1}{\bar{\rho}(1+w)}\qty(\hat{k}_i\hat{k}_j-\frac{1}{3}\delta_{ij})\Sigma^{ij}~.
\end{equation}
This quantity is also gauge invariant, but neither Einstein's nor conservation equations specify its evolution.

Hu originally motivated the closure relation choice to recover a neutrino-like fluid behavior in the appropriate limit. The choice from \cite{Kopp} maintains this behavior with a slight change to allow the crossing of $w=0$. In our notation, this relation becomes;

\begin{align}
    \dot\sigma_g=-3\mathcal{H}\sigma+\frac{8k^2}{3(1+w)}c_v^2\hat\Theta_g~, \\
    \text{with } \hat\Theta= \theta/k^2-\zeta-\frac{1}{2}\dot\nu~, \nonumber
\end{align}

\noindent where $\nu$ and $\zeta$ are scalar perturbations of the metric\footnote{These are often also denoted as $B \equiv \zeta$ (scalar potential of time shift vector) or $E \equiv H_L \equiv \nu$ (scalar potential for the traceless part of the spatial metric). \cite{Malik:2008im,Durrer:2004fx,Brandenberger:2003vk,baumann_2022,riotto2014lecture}}, contributing as the time-space component $-\grad_i\zeta$ and the traceless space-space component $\qty(\grad_i\grad_j-\frac{1}{3}\gamma_{ij}\vec{\grad}^2)\nu$.  The variable $c_v^2$ is a free parameter of the model and is similar to shear viscosity in fluid dynamics; see section IV.A of \cite{Kopp} for more details on this connection. The gauge invariant variable $\hat\Theta$ reduces to the velocity perturbation in the conformal Newtonian gauge where the last two terms vanish. One recovers free-streaming massless neutrinos truncated at the quadrupolar moment by setting $w=c_s^2=c_v^2=\frac{1}{3}$.

In full generality, the parameters of GDM can take a functional form of $w(a), c_s^2(a,k)$, and $c_v^2(a,k)$. However, our analysis restricts the latter two to constants, given that the freedom in $w(a)$ is already sufficient to draw conclusions about the exploration of such high-dimensional models.

\subsection{Numerical implementation}

We now discuss our implementation of our model of GDM via modification of the publicly available Einstein-Boltzmann solver \class{} \cite{classII}.

First, we discuss our choice of parameters to describe the model.
Traditionally energy density is set by the fraction of energy density today $\Omega_{\rm g}$\,. However, the flexibility in the history of $w(z)$  can make setting an appropriate range difficult. Instead, we pick a redshift of interest $z_{\rm g}$  at which we set the fractional contribution of GDM $f_{\rm gdm}\equiv\rho_{\rm g}(z_{\rm g})/\qty(\rho_{\rm r}(z_{\rm g})+\rho_{m}(z_{\rm g})+\rho_{\rm g}(z_{\rm g}))$. When $z_{\rm g}$ is large enough so that dark energy does not play a significant role, $f_{\rm gdm}$ is a good approximation of the fraction of the energy density that GDM provides at $z_{\rm g}$\,. To set the energy density of dark energy (in our case in the form of a cosmological constant), we integrate $\rho_{\rm g}$ forward until today before using curvature constraints to set $\Omega_\Lambda$.
 
The history of $w(z)$ is determined by a set of $N$ pairs of control points $(\log_{10}(a_1),w_1)\ldots(log_{10}(a_N),w_N)$. Between points, piece-wise linear interpolation occurs. We used logarithmic space rather than linear to give finer control in any decade. This interpolation is a simple way to describe a continuous function without needing additional guards to enforce the physical requirement that $-1\leq w\leq1$ even between control points, which can plague smoother interpolation schemes.\footnote{The largest drawback in our choice is that $c_a^2$ will not be continuous as it involves $\dot{w}$. If continuity of $c_a^2$ were important for some application, other choices are possible such as a cubic piece-wise monotonic spline which enforces monotonicity between control points.}

Valid inputs of control points must range from when \class{} initializes the background, $a_{\rm ini}=10^{-14}$ by default, to today. To ensure radiation domination for initial conditions, we set the restriction that $w\leq1/3$ at early times. Appendix A derives the initial conditions. We denote this realization of GDM with complete freedom over the $w(a)$ evolution through a multitude of control points as high-dimensional GDM (HDGDM).

At points within \class{}, we must choose a decomposition of energy density sources into radiation and pressureless matter.\footnote{This is important, for example, when \class{} sets BBN-consistent values of the Helium abundance, or computes small corrections to the initial conditions. The overall impact on observables is rather minor, though.} However, there is no singular way to divide $\rho_g$ into materials of a constant equation of state parameter. Our choice splits GDM into two materials depending on the value of $w$. We divide up the range of $w$ into intervals with endpoints  $(-1,0,1/3,1)$ corresponding to dark energy, non-relativistic matter, radiation, and a free scalar field, respectively. For a $w$ contained in interval $(w_I,w_J)$, we divide $\rho_{\rm g}$ into
\begin{equation}
    \rho_{I,{\rm g}} =\frac{w-w_J}{w_I-w_J} \rho_{\rm g} \text{ and } \rho_{J,{\rm g}} =  \rho_{\rm g}-  \rho_{I,\rm g}~.
\end{equation}
This decomposition ensures that each component is non-negative and is continuous in the transition between intervals.\footnote{This decomposition also recovers the usual neutrino splitting as $\rho_r \approx 3P_\nu$ and $\rho_m \approx \rho_\nu-3P_\nu$ that is used in \class{}.}

As described above, $c_s^2$ and $c_v^2$ are additional free parameters affecting the evolution of the perturbations. Both perturbation parameters lie in the range $[0,1]$ in order to support growing modes and to have characteristic scales grow at subluminal rates. Our current implementation has these parameters as constants in time. By default, if the user does not set $c_s^2$, then non-adiabatic pressure (NAP) is turned off, forcing $c_s^2=c_a^2$. 

\subsection{Numerical validation of the GDM implementation}

Generalized dark matter is a very general material that can mimic many other materials already implemented in \class{}. We first compare against perfect fluids of constant equation of state parameter less than 0 before looking at one that linearly varies with scale factor. Subsequently, additional neutrinos and ``self-interacting dark radiation" (SIDR)  with $w=1/3$ with and without free streaming, respectively, are modeled. Our \class{} implementation can reproduce the temperature, polarization and lensing (TT, TE, EE, $\phi\phi$) power spectra for all the test cases discussed below to within a precision goal of $0.1\%$ of cosmic variance for $\ell < 4000$, sufficient for our applications.

\class{}  already has an implementation of a non-perfect fluid (\texttt{fl}). It must have a time-varying equation of state parameter of the from $w_{\texttt{fl}}(a)=w_0+w_a(1-a/a_0)$ under the restrictions that $-1<w(a_\mathrm{ini})<0$ (with $w(a_\mathrm{ini}) \approx w_0+w_a$). These fluids serve as a helpful place to start comparisons. 

We first validate with $w_{\texttt{fl}}(a) = w_0$ (explicitly setting $w_a = 0$). For the HDGDM we set $w=w_0$ at initial and final times with no viscosity or NAP. To stay reasonably close to the observed data, we set $\Omega_{\rm g} = \Omega_{\texttt{fl}}=0.1$ while keeping all other parameters to their default values in \class. Due to the strict bounds for $w_0+w_a$ we compare against $w_0=\{-1/10,-1/3,-2/3,-9/10\}$. The resulting spectra agree to within our 0.1\% goal. 

By allowing $w_a$ to be non-zero, we can make an additional comparison. Due to the different functional forms for time dependence, $\texttt{fl}$'s linear function in $a$ can only be approximately described by the piece-wise linear $w(\log a)$ of HDGDM. We use $249$ control points with increasing density as $a$ tends to 1 and manage to achieve our target accuracy. We compare these fluids at all combinations of $w(a_\mathrm{ini})$ and $w_0$ drawn from $\{-1/10,-1/3,-2/3,-9/10\}$ with all other parameters treated the same as above, and find excellent agreement.

Comparison for materials with $w(a_{\rm ini})>0$ can also be made by attempting to mimic the behaviors of neutrinos or self-interacting dark radiation (SIDR). By design, the shear stress of GDM recovers the same behavior as the truncated neutrino hierarchy when $c_v^2=1/3$. For the best comparison, we set the fluid approximation of \class{} to match Hu's formalism \cite{Hu:1998kj}. To set the energy contribution, we contrast between an increase in $N_{\rm ur}$ by $10\%$ and the equivalent $\Omega_g$ while setting $w=1/3$ for all time. Similarly, by turning $c_v^2$ off and replacing the additional $N_{\rm ur}$ with $N_{\rm idr}$, we can subsequently reproduce the spectra of SIDR within our precision goal.

We conclude that our realization of GDM in \class{} allows for a robust modeling of a wide variety of equations of state to high precision. While execution times of specialized materials can outperform in their specific domains, the ability to seamlessly transition from one material to another has otherwise been impossible. Subsequently, we leverage this capability to search for models satisfying cosmological data while bringing concordance to the observed tensions.

\section{Model Space Exploration}

The traditional approach for exploring the implications of a model space and a given data set is to calculate the posterior probability distribution of the model parameters given the data and visually examine one- and two-dimensional marginal posterior distributions. Here we emphasize that applying such methodology to relatively high-dimensional spaces can lead to unusual and undesirable results. For example, one might naively think that broadening a model space by introducing additional parametric degrees of freedom or relaxing priors on these degrees of freedom, would necessarily lead to broader distributions of parameters of interest, such as the marginal posterior distribution of $H_0$. The breadth of the prior, and in particular its level of support for $H_0$ values of, say, 73 km/s/Mpc, would then indicate how well the model space in question can accommodate a high $H_0$ value. We have found, though, that this is not always the case. In fact, we have found that narrowing the parameter space of a model to a subspace that contains regions that admit good fits to the combined CMB and SH0ES data can nevertheless lead to a {\em wider} marginal $H_0$ distribution compared to the original. This effect has been noted in lower-dimensional cases before, such as in \cite{Smith:2020rxx,Schoneberg:2022grr} where it had a marginal impact. However, in the high dimensional broad HDGDM parameter space this prior volume effect is qualitatively more important (see Fig.~\ref{fig:d-triangle}).

This section presents a worked example in which we can observe and understand the cause of this unusual feature of the traditional investigation. The emergence of this undesirable result -- of interesting solutions becoming effectively lost -- leads us to examine how the traditional approach is not suited for us, given our particular goals for exploring the implications of data. Although our primary goal in this section, and with this paper, is to point out the pitfalls of high-dimensional model space exploration, we briefly discuss what one can do instead. Finally, we include some takeaways from our investigations that we can manage to make despite the pitfalls we are emphasizing.

\subsection{Standard Techniques}

We begin using HDGDM to explore a parameter space we knew beforehand contained an interesting solution to the $H_0$ tension. In particular, we choose as a motivating model ``Wess-Zumino Dark Radiation" (WZDR) explored in \cite{Aloni:2021eaq,Schoneberg:2022grr,Allali:2023zbi}. 

Models like WZDR and potentially other near radiation-like fluids introduce light particles with a decay channel into massless particles. The combined system of particles starts with an equation of state $w=1/3$, which decreases as the particles' rest mass energy becomes significant compared to the thermal energy of the particles, after which only the massless species remains and the equation of state returns to $w=1/3$.

Our toy version of these models is HDGDM with $w=1/3$ for times before $a=10^{-5}$ and after $a=10^{-2}$. We include five equation-of-state parameters $\{w_1, w_2, \ldots w_5\}$ at scale factors $\{10^{-4.5}, 10^{-4}, \ldots 10^{-2.5}\}$ respectively. For a preliminary investigation, we turn off non-adiabatic pressure (NAP) and viscosity and set uniform priors over $[-1,1]$ for $w_i$ and $[0,0.7]$ in $f_{\rm gdm}$ at $z_{\rm g}=3000$. We call this model ``Step-like" (SL). 

For all models explored and the baseline \LCDM\ model, we adopt flat priors over standard cosmological parameters $\{H_0,\omega_b,\omega_c,\ln(10^{10} A_\mathrm{s}), n_s, z_{\rm reio} \}$. The neutrino sector assumes one massive, $m_\nu=0.06 {\rm eV}$, and two massless species (following \cite{Planck:2018vyg}). We use the ``halofit" module of \class{} to include non-linear effects.

Following standard approaches for model evaluation, we produce the posterior distribution via the Polychord algorithm inside of Cobaya \cite{cobaya, polychord}. We combine several data sets into the collection, $\mathcal{D}$, which includes data from Planck, BAO measurements, and Pantheon. In particular, we make use of Planck 2018 TT, TE, EE for low and high $\ell$ modes (marginalized over the nuisance parameters) as well as the lensing reconstruction \cite{Planck2018paramest,Planck2019GravLens}. BAO measurement data is taken from BOSS DR12 \cite{Mueller:2016kpu}, 6dF \cite{Beutler:2011hx} and MGS \cite{Ross:2014qpa}. Lastly, we include Pantheon supernova data \cite{Pan-STARRS1:2017jku}. 

\begin{figure}
\includegraphics[width=0.9\linewidth]{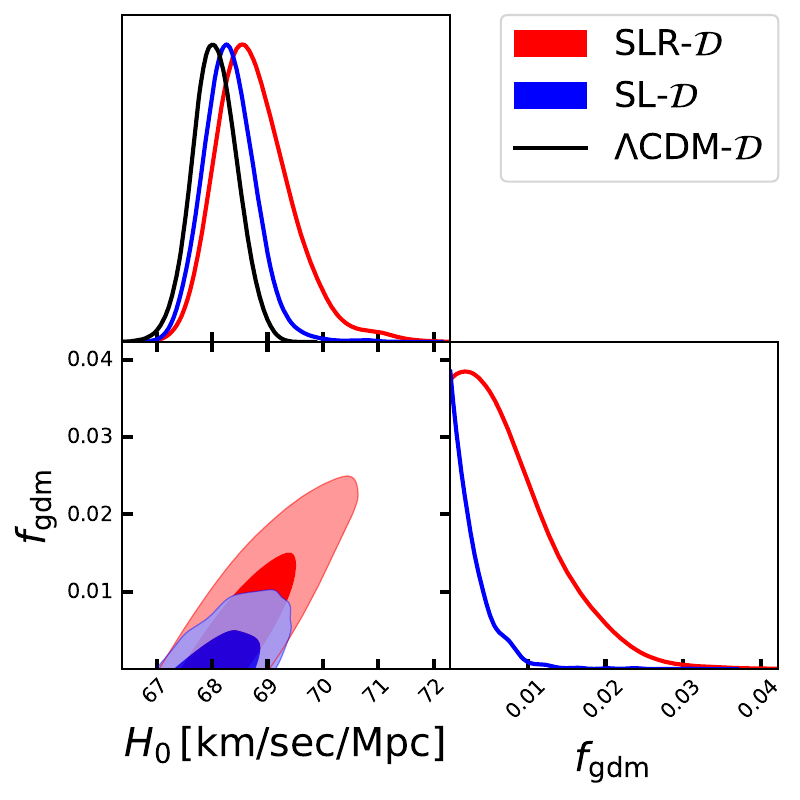}
\caption{\label{fig:d-triangle} Triangle contour plot of $H_0$ and $f_{\rm gdm}$ for SL and SLR with $\mathcal{D}$. The shaded regions indicate the 68\% and 95\% credible regions.}
\end{figure}

We show some of the posteriors of the SL model in Fig.~\ref{fig:d-triangle}. One might initially conclude that there is minimal support for substantial $f_{\rm gdm}$ values and, by extension, GDM; however, this conclusion would be a questionable one as the result is highly dependent on the implicitly assumed priors (as we will soon see) and the priors are chosen in some sense arbitrarily. The posterior that we calculate is only the posterior {\em given} these priors: that the model is correct (at some point in the parameter space) and that the probability (prior to examining any data) is distributed uniformly across the parameter space. These priors do not reflect our true prior beliefs about this model space, one we are only adopting for phenomenological exploration.

\begin{figure*}
\includegraphics[width=\linewidth]{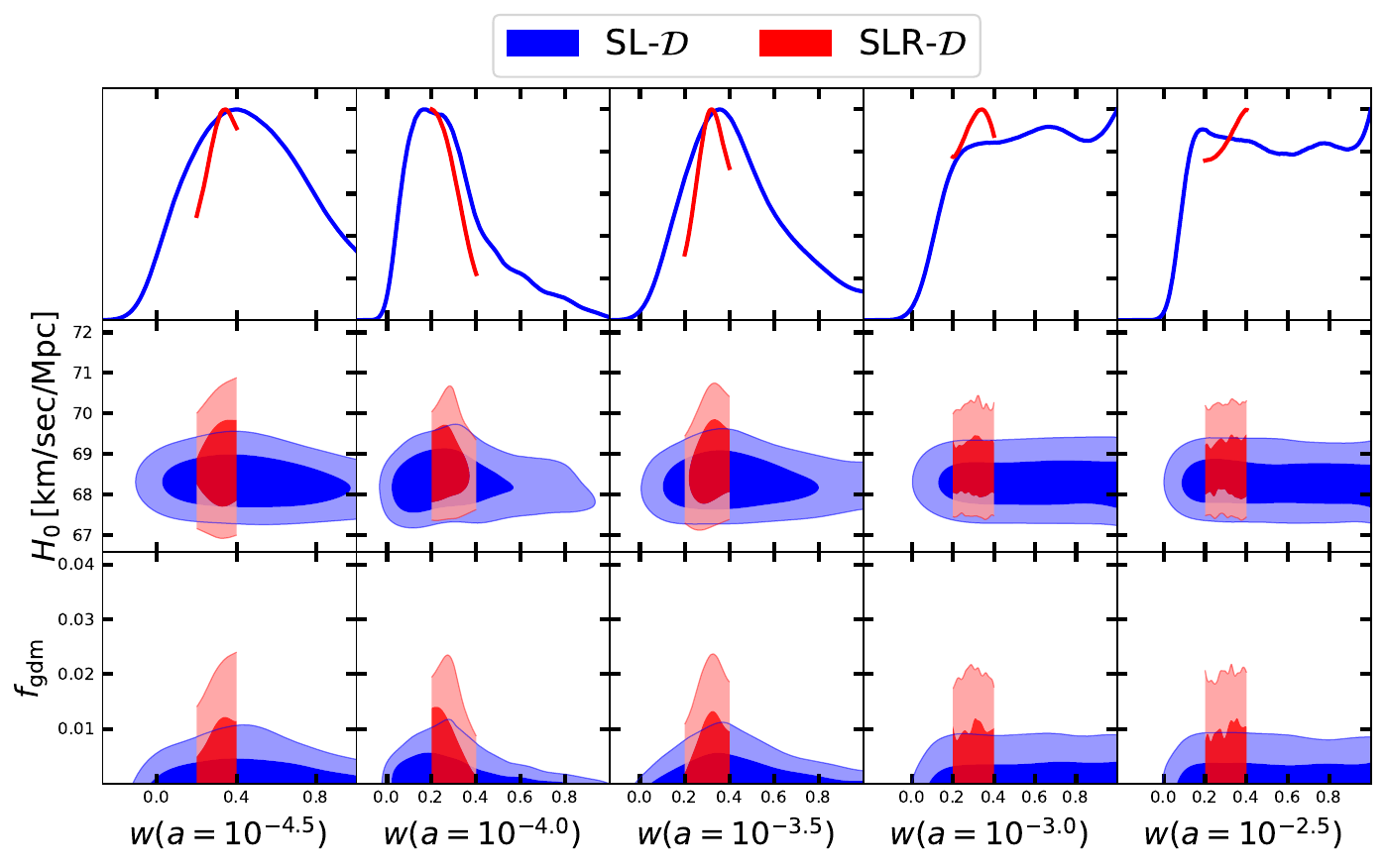}
\caption{\label{fig:d-rectangle} The posterior of each $w_i$ parameter of SL and SLR models along with their contours in relation to $H_0$ and $f_{\rm gdm}$ using the $\mathcal{D}$ data collection.  The shaded regions indicate the 68\% and 95\% credible regions.}
\end{figure*}

We show the marginalized posteriors of the $w_i$ in Fig.~\ref{fig:d-rectangle}. Here we observe a band of values in the earliest three $w_i$\,, which both support larger $f_{\rm gdm}$ and coincide with preferred values of $w_i$\,. This observation motivates a ``Step-like restricted"  model (SLR) where $w_i$ are all restricted to a uniform prior from $[0.2,0.4]$. This choice narrows SL closer to behaviors seen in models like WZDR, where we expect typical $w$ values to fall in the $[0.25,0.33]$ range. Figures \ref{fig:d-triangle} and \ref{fig:d-rectangle} show the resulting posterior in terms of $H_0$ and the model parameters.

We see much higher values of $f_{\rm gdm}$ now contained in the 95\% credible region. More interesting is the impact on the distribution of $H_0$\,. While we expect an upward shift in peak probability, given the increased amount of the HDGDM, the broadening of the distribution is a surprise. When going from SL to SLR, we have narrowed our priors of each $w_i$\, and thus narrowed the total prior parameter space (reducing the prior uncertainty). This restriction has effectively added information, yet we see less certainty in the posterior of $H_0$\,. 

We attribute the change in predictions to a volume effect. In SL, when $f_{\rm gdm}$ tends to zero, any impact of GDM is minimal, reducing the model to effectively \lcdm. As a result, the marginalized distribution for $H_0$ gets uniform contributions from the whole of the $w_i$ parameter space. As we increase $f_{\rm gdm}$, only the $w(z)$ histories producing outputs that fit data well find significant weight. Naturally, these permitted histories take up a small portion of all possible values in $w_i$\,. In SLR, we have reduced the $w_i$ parameter space's total volume increasing the ratio of supported volume to prior volume. Thus the volume effect has a diminished boost when $f_{\rm gdm}\approx 0$. Furthermore, the remaining region contains histories that can fit data well, allowing $f_{\rm gdm}$ to find support above zero. Thus, we see that looking primarily at the posterior region of support might cause premature conclusions on the existence of viable regions within the SL model away from $\Lambda$CDM.

The failure of standard inference methods to distinctly pick out the potentially interesting area of model space raises the question if it is the correct choice for our purposes. When conducting parameter estimation, using marginalized distributions is a natural choice. In contrast, our observations in this section call for alternative methods when seeking a region to focus on in large model spaces which can be sensitive to prior assumptions. We now review a few of those techniques.

\subsection{Model Searching Techniques}
The peculiarities between SL and SLR cue us to examine our goals and methods. The fact that we are searching for a sub-model rather than estimating parameters is critical here. Marginalized posteriors, vital to the latter, can mislead us in the former. We briefly discuss other tools helpful in finding new models of interest. To begin, we present the technique we used (minimizing the log-likelihood via simulated annealing) before suggesting potential alternatives for other problems. 

Approaches which look for parameters that provide good fits to data are particularly relevant to our goals. We can easily discard a model that has no improvement in quality of fit when compared to \lcdm. If a notable region can provide good fits, then features of that region could be extracted to motivate theoretical models. Thus, relevant techniques must find the existence and location of these regions.

An important tool to discover potential subspaces or validate choices made for other reasons is $\chi^2$ minimization. However, the curse of dimensionality of the HDGDM model has proven to make such a search challenging. Even our relatively small dimensional extension SL has many local minima. Additionally, parameters like the last two $w_i$ parameters and their near-flat posterior may also have weak (and potentially noisy) relations to data. Either of these problems may cause standard minimizing algorithms to have issues. We have found that simulated annealing has had the most success though not without the need for care.

In practice, we start searches with several independent Markov chains, which may originate in widely separated initial points. We then seed a second collection of chains with initialized parameters drawn from a normal distribution centered at the best-fit of the first set. We examine the second collection of chains to ensure no wide swings in the best-fit parameter values. Interestingly, most of our primary collection of SL chains ends inside the SLR region, with the overall best-fit of the secondary run not far from this region. 

More advanced minimization techniques based on annealing are currently being developed, and can be compared to other robust minimization approaches such~as~\cite{2018arXiv180400154C,2018arXiv181211343C}.\footnote{We note that simplex-based methods like \cite{10.1093/comjnl/7.4.308} would typically be far too slow in such high-dimensional parameter spaces}

Although we do not use it here, likelihood profiling, as done for example in \cite{Herold:2021ksg}, can also be a helpful tool to see how the best fit of a model evolves as we vary a parameter of interest. Given that there is no prior weighting in likelihood profiling, this will naturally outline parameter regions of good fit, as long as good convergence of the profiler is ensured.

Another model search technique returns to MCMC. One could use the result of these chains with some threshold cut-off in log-likelihood to locate localized regions of parameter space which provide good quality fits. In contrast to traditional techniques, the focus on the quality of fit rather than the marginalized posterior will get around potential volume effects. This search could be aided by changing the temperature of the MCMC as well; by making the chain run colder one would further emphasize regions of good fit.

\begin{figure}[t]
\includegraphics[width=0.9\linewidth]{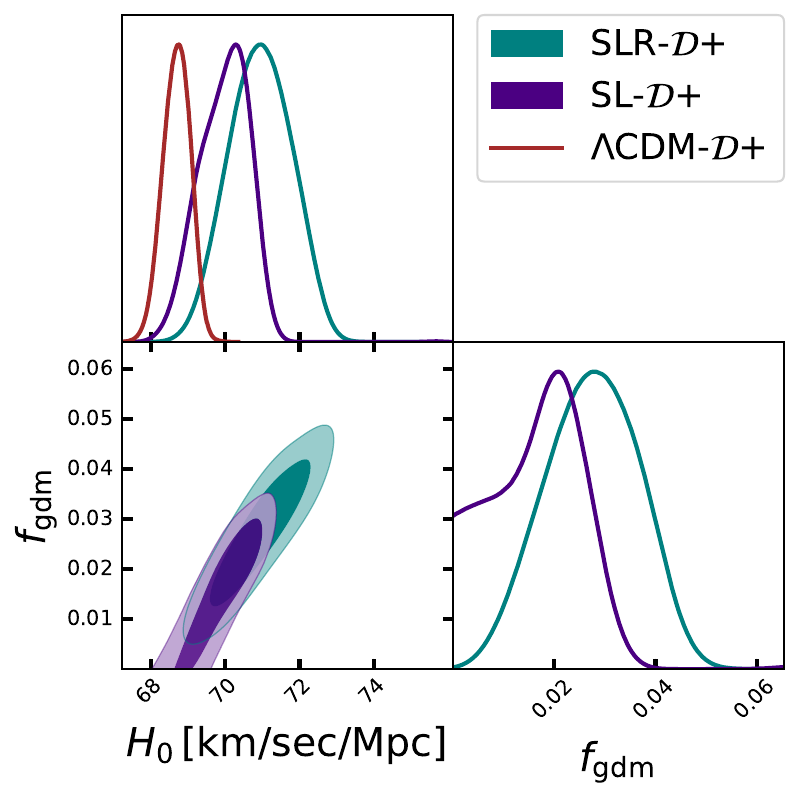}
\caption{\label{fig:d+-triangle} Triangle contour plot of $H_0$ and $f_{\rm gdm}$ for SL and SLR with $\mathcal{D}+$. The shaded regions indicate the 68\% and 95\% credible regions.}
\end{figure}
\begin{figure*}[t]
 \includegraphics[width=\linewidth]{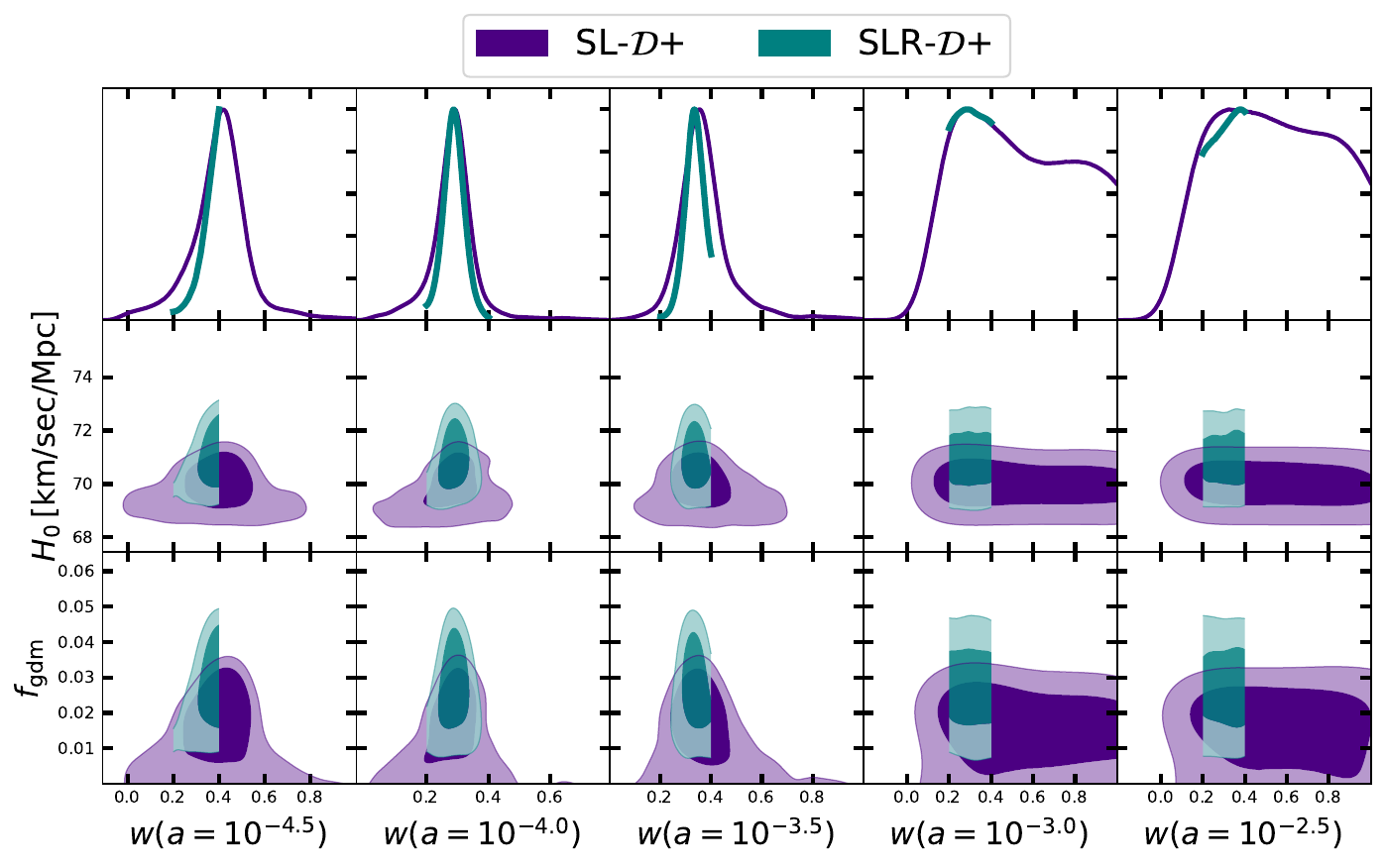}
\caption{\label{fig:d+-rectangle} The posterior of each $w_i$ parameter of SL and SLR models along with their contours in relation to $H_0$ and $f_{\rm gdm}$ using the $\mathcal{D}+$ data collection. The shaded regions indicate the 68\% and 95\% credible regions.}
\end{figure*}

\subsection{Hubble constant prior}

In this subsection, we extract additional insights from SLR, including its flexibility when attempting to fit additional data and what difference we see from our original motivating model. We will see that solutions exist within the model that bring concordance across CMB anisotropies, the Cepheid and supernova based distance ladder, and BAO data. 

We first explore an expanded data set collection, $\mathcal{D}+$, which adds SH0ES data to our original collection \cite{Riess:2021jrx}. We see some preliminary results in Fig. \ref{fig:d+-triangle} and Fig. \ref{fig:d+-rectangle}. Adding SH0ES lowers the quality of fit \LCDM\ can provide while $f_{\rm gdm}$ is small. Thus, the volume effect boost is diminished even in the SL model. Furthermore, once we move from SL to SLR,  there is a clear preference for $f_{\rm gdm}>0$, 
with $f_\mathrm{gdm}$ shifting from $f_\mathrm{gdm} < 0.030$ (95\%CL) to $f_\mathrm{gdm} = 0.028\pm 0.0098$, and a corresponding 1-$\sigma$ shift in $H_0$ from $70.0\pm 0.7$ km/s/Mpc to $70.9\pm 0.8$ km/s/Mpc. These constraints can also be compared to the posterior without the addition of the SH0ES prior, which is at only $H_0 = 68.3 \pm 0.5$ km/s/Mpc for the SL model.
An interesting consideration to draw from this is that in order to check the validity of a given model in terms of easing a particular tension, it can be extremely helpful to force it into the beneficial parameter space by including the additional data.

We note that under $\mathcal{D} +$, the posterior of $w_1$ at $a=10^{-4.5}$ is cut off by the SLR prior. However, even given an extended range up to 0.6 in $w_1$\, the properties of the model are not significantly different, so for brevity, we do not introduce another model. However, this may hint that something of interest lies in the $w>1/3$ range, a region inaccessible to most light particle models. 

To get a better understanding of the properties of the models, we examine their respective best-fits. In our explorations of these spaces, all sampled parameters have flat priors except for $A_{\rm planck}$ of the Planck data set, which has a normal distribution. Thus for a fair comparison, we consider the sum of $\Delta \chi^2$ for each data set along with twice the negative of the log prior of $A_{\rm planck}$, discarding any other priors as they provide only constant effect within their respective models. The results can be seen in Table \ref{tab:chi2-comp}, where it becomes clear that despite the smaller space, the best-fit of the restricted model is very close in quality of the best-fit over the larger space. In contrast to \LCDM\, the search demonstrates that better fits to SH0ES data exist while improving the fit to Planck data. The most significant pushback to the new parameters comes not from Planck but from BAO data.

\begin{table*}
    \begin{tabular}{ |c|c||c|c|c|c|c||c||c|c|c|c| } 
    \hline
    Data set      & Model & $\Delta \chi^2_{\rm CMB}$ & $\Delta \chi^2_{\rm BAO}$ 
     & $\Delta \chi^2_{\rm SN}$ & $\Delta \chi^2_{\rm SH0ES}$ & $-\Delta 2\ln \pi(A_{\rm planck})$ & Total & $H_0$ (Km/s/Mpc) & $\Delta H_0$ & $ S_8$ & $\Delta S_8$ \\ 
     \hline\hline
     $\mathcal{D}$ & SL   & -1.33 & 0.20 & -0.01 &        & -0.05 & -1.19 & 68.38 & 0.32 & 0.837 & 0.007\\
     \hline
     $\mathcal{D}$ & SLR  & -1.32 & 0.20 & -0.02 &        &  0.03 & -1.11 & 68.67 & 0.56 & 0.836 & 0.006\\ 
     \hline
     $\mathcal{D}+$ & SL  & -0.74 & 0.45 & 0.00     & -15.67 & -0.11 & -16.07 & 71.66 & 2.96 & 0.829 & 0.012 \\
     \hline
     $\mathcal{D}+$ & SLR & -0.38 & 1.07 & 0.02  & -16.20 & -0.15 & -15.63 & 71.88 & 3.18 & 0.829 & 0.012 \\
     \hline
    \end{tabular}
    \caption{\label{tab:chi2-comp} Change in $\chi^2$ values in comparison to the \LCDM\ best-fit of the same data set. We also show the shift in the $H_0$ and $S_8$ bestfit values.} 
\end{table*}
This better fit can also be seen through the $Q_\mathrm{DMAP}$ criterion introduced in \cite{Raveri:2018wln}. In our case, we obtain a $Q_\mathrm{DMAP}$-based Hubble tension of $4.7\sigma$ in $\Lambda$CDM, while in SL we obtain only $2.6\sigma$ and in SLR $2.7\sigma$. These $Q_\mathrm{DMAP}$ show (by definition) the expected ordering that the models with more restricted parameter spaces perform worse. In particular, the ability of the HDGDM model to ease the Hubble tension in some part of the large parameter space becomes far more evident using $Q_\mathrm{DMAP}$ as the metric and is not overshadowed by prior volume effects.\footnote{This, of course, does not imply that the HDGDM model is a particularly useful model, based on its mostly phenomenological construction. Indeed, its Aikaike information criterion $\chi^2+2N_\mathrm{param}$ shows an overall difference of 10.9 without the Hubble prior (and -4 with the Hubble prior) due to the large parameter penalty, thus being easily outperformed by other smaller models. However, it does imply that the HDGDM can indeed be used as a valuable tool to aid in the search for mechanisms of reconciling the tension.} Thus the $Q_\mathrm{DMAP}$ criterion can be used as a fast and relatively cheap check to see if further exploration of the model parameter space might be sensible.

\FloatBarrier

\begin{figure*}\includegraphics[width=0.9\linewidth]{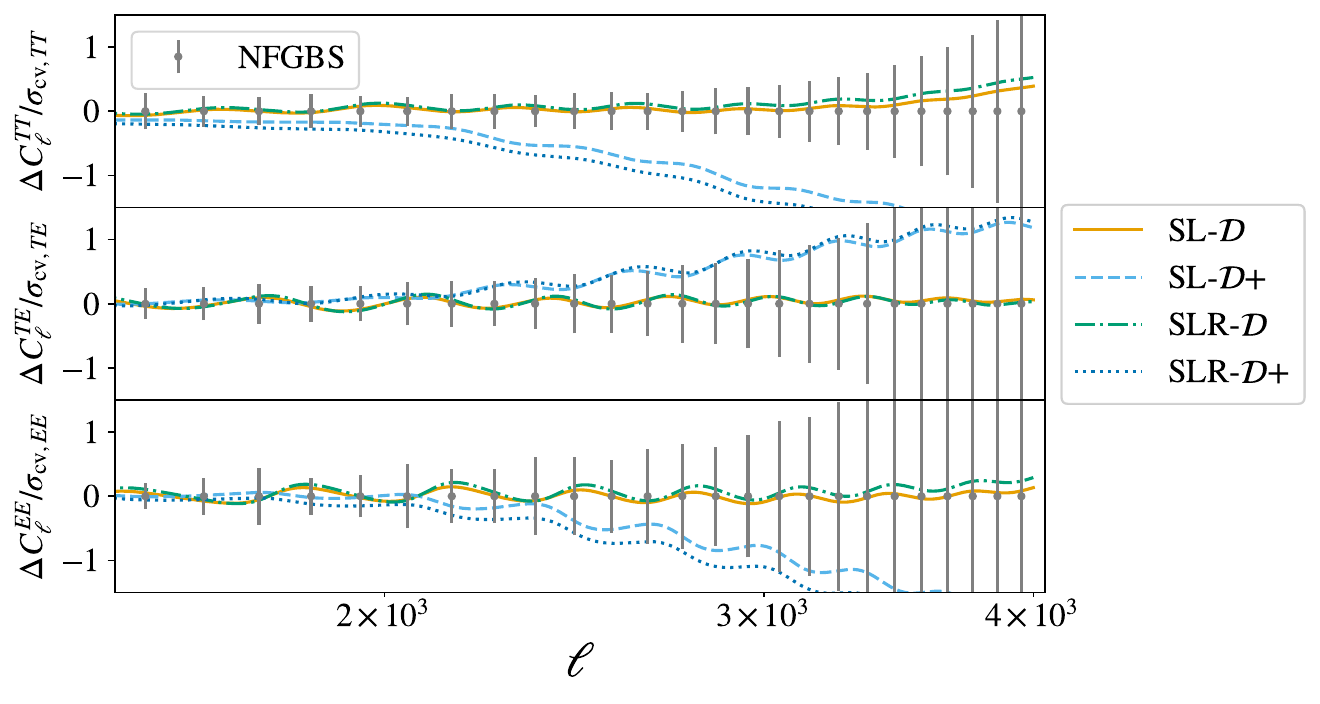}
\caption{\label{fig:deltacl-ttteee} Best-fit power spectra for each model and data collection, with the fiducial \LCDM best-fit to $\mathcal{D}$ spectrum subtracted, divided by the cosmic variance of the same fiducial. The error bars are a forecast of near-future ground based survey.
}
\end{figure*}

\begin{figure}
\centering
\includegraphics[width=\linewidth]{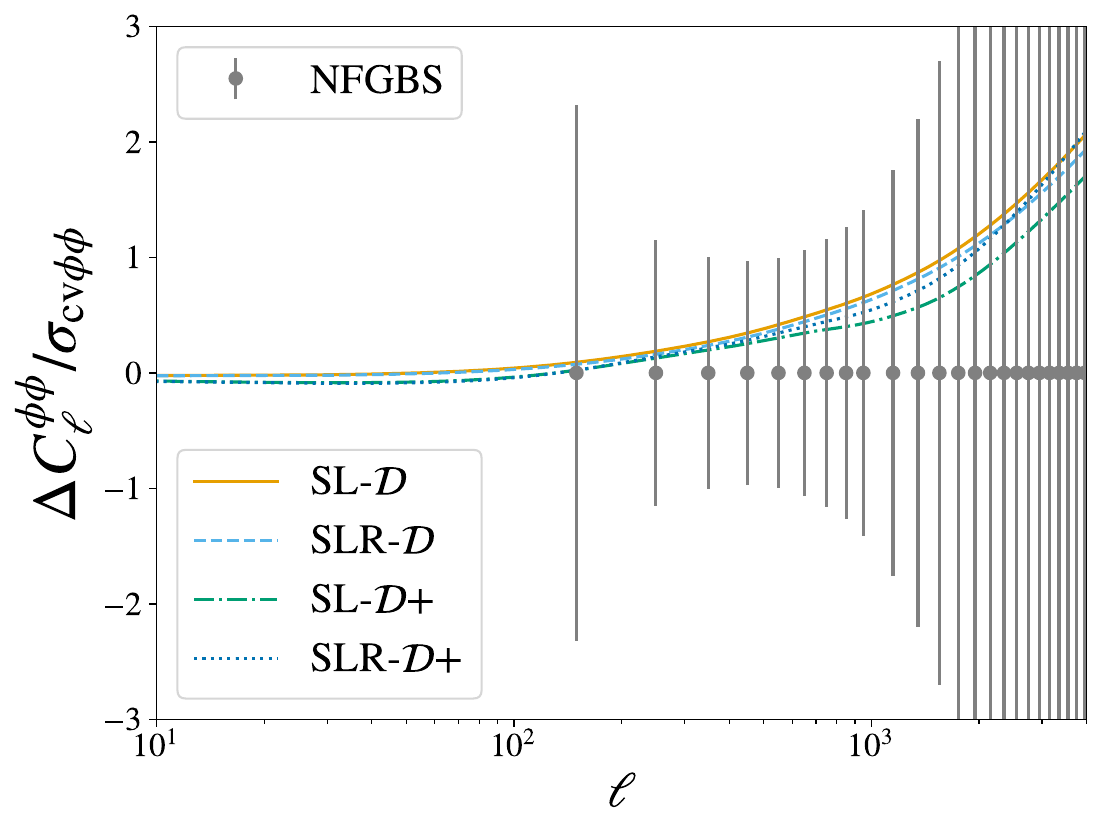}
\caption{\label{fig:deltacl-pp} Best-fit lesning ($\phi\phi$) power spectra for each model and data collection, with the fiducial \LCDM best-fit to $\mathcal{D}$ spectrum subtracted, divided by the cosmic variance of the same fiducial. The error bars are a forecast of near-future ground based survey}
\end{figure}
 
\subsection{Noteworthy features}
An ideal outcome of this work would be to inspire the creation of a physical model with features found in our SL model, which can make predictions that are distinguishable from \LCDM\ using future observations. We begin by considering the spectra of the best-fit parameters to establish how they differ from \LCDM. Subsequently, we consider what typical histories of $w(z)$ and $H(z)$ a physical SL model might contain. We finally consider if there are any implications for $S_8$ tensions. 

We compare the best-fits of SL and SLR to \LCDM, in Fig. \ref{fig:deltacl-ttteee} and Fig.\ref{fig:deltacl-pp}. In these plots, we compare the change of the respective spectra as a fraction of cosmic variance. For a relevant scale we provide forecasted error bars of a 10\,000 square degree survey reaching a depth of 9 $\mu$K-arcmin in temperature and 12 $\mu$K-arcmin in polarization, which we refer to as NFGBS (near-future ground based survey). Such a survey could be conducted in the next few years with existing instruments, such as SPT-3G \cite{SPT-3G:2021vps}.

Without SH0ES data, the SL and SLR best-fit spectra differ very little from the \lcdm\ best-fit. The most significant difference is increased power of the $\phi\phi$ and TT spectra at small scales\footnote{We note that this increase in lensing power is primarily driven by differences in the $\Omega_m h^2$ bestfit from the CMB anisotropy data allowed by the additional HDGDM, while the direct impact of the HDGDM on the CMB lensing is instead very minute.}.

With SH0ES, the effect on the TT, TE, EE spectra becomes more pronounced, causing changes in TT that should be detectable with upcoming observations. All models considered fit the current spectral data well but have potentially divergent behaviors at $\ell$ values beyond current data. We consider the ability of NFGBS for $\ell\leq4000$ to distinguish our models from $\Lambda$CDM. The combined signal to noise ratios of the (TT, TE, EE, $\phi\phi$) spectra are found to be 2.5 for SL with $\mathcal{D}$ (2.6 for SLR) and 11.0 with $\mathcal{D}+$ for SL (13.5 for SLR) with the lensing reconstruction being the largest contributor to former and the temperature anisotropies overtaking for the latter. We thus expect near-future data to be capable of discriminating between the best-fit solutions of \LCDM{} and SLR over the $\mathcal{D}+$ data collection.

\begin{figure*}
\centering
\includegraphics[width=0.9\linewidth]{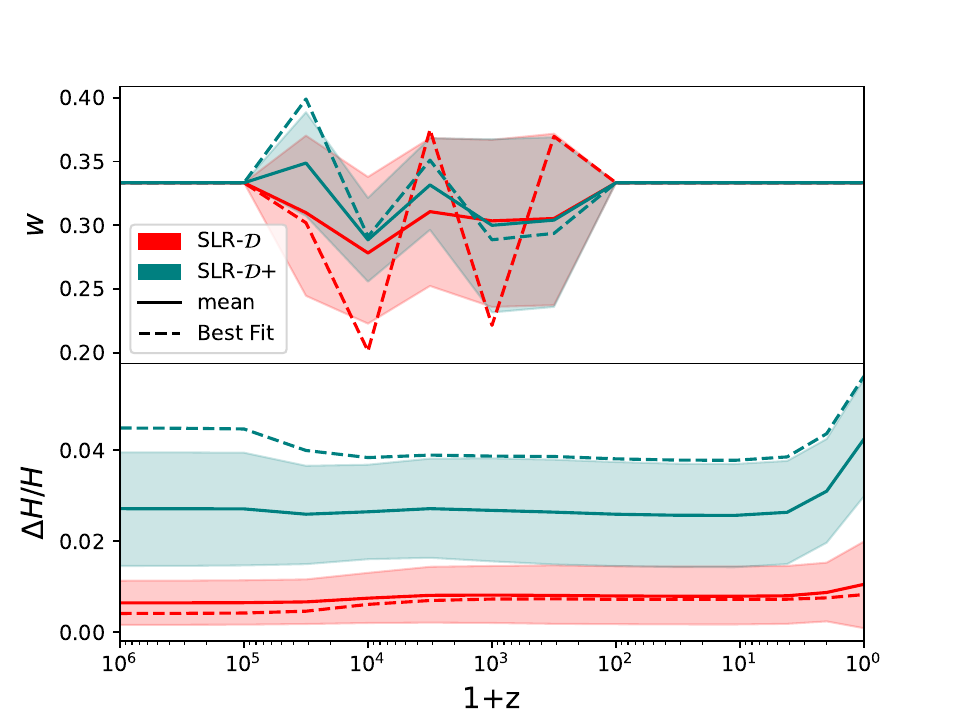}
\caption{\label{fig:SLR-history} The history of $w(z)$ and the fractional change in the history of $H(z)$ from the fiducial \LCDM\ best-fit for the SLR model. The shaded region indicates the 68\% credible region.}
\end{figure*}

In Fig. \ref{fig:SLR-history}, we show preferred histories of $w(z)$ and $H(z)$. The plot presents the Hubble parameter as a fractional change from the fiducial values of the best-fit value of \LCDM\ on the $\mathcal{D}$ data collection. 

Overall the $\mathcal{D}$ dataset does not constrain $w(z)$ particularly well, with only a minor trend to lower $w(z)$ at $z\sim10^4$ and a return upwards for the next node at $z\sim10^{3.5}$. Nevertheless, the mean shape does resemble the behaviors of \cite{Aloni:2021eaq}. With the addition of SH0ES data $f_{\rm gdm}$ is larger and thus HDGDM has an increased impact. Thus, the history of $w(z)$ have more impact on observable. The three control points before $z=10^3$ become more constrained, with the latter two points about as free as before. These constraints reinforce how the epoch leading up to matter radiation equality and recombination can most impact predictions and, consequently, how sensitive data are to these parameters. Particularly interesting are the features of a large spike of $w>1/3$ at $z \sim 10^{4.5}$ and a dip at $z \sim 10^4$. While the latter is known in the context of WZDR, the former is quite interesting as a starting point for potential future exploration of mechanisms easing the Hubble tension. We leave a more detailed investigation of this spike feature to future work. Even when focusing on the mean behavior, we see that SLR can make a significant shift of $H_0$ without extreme spikes or dips, supporting the idea that the restricted model space still has more than enough flexibility to fit $\mathcal{D}$ data while relieving tension with SH0ES observations. In contrast, even with the freedom that HDGDM provides, the Hubble expansion histories do not seem to broaden significantly, with the $1\sigma$ upper bound of the SLR case only being about $1\%$ (resp. $4\%$) larger than in the \LCDM\ fiducial for $\mathcal{D}$ (resp. $\mathcal{D}+$). However, as before, these bounds should not be misinterpreted. They represent the marginalized preference of histories of $H(z)$ for a given model with particular priors, but are not to be seen as solely data-driven bounds on possible deviations in expansion histories. Indeed, this is why the contours for the two models can disagree for the same dataset at the $\sim 1\sigma$ level.

\begin{figure}
\centering
\includegraphics[width=0.9\linewidth]{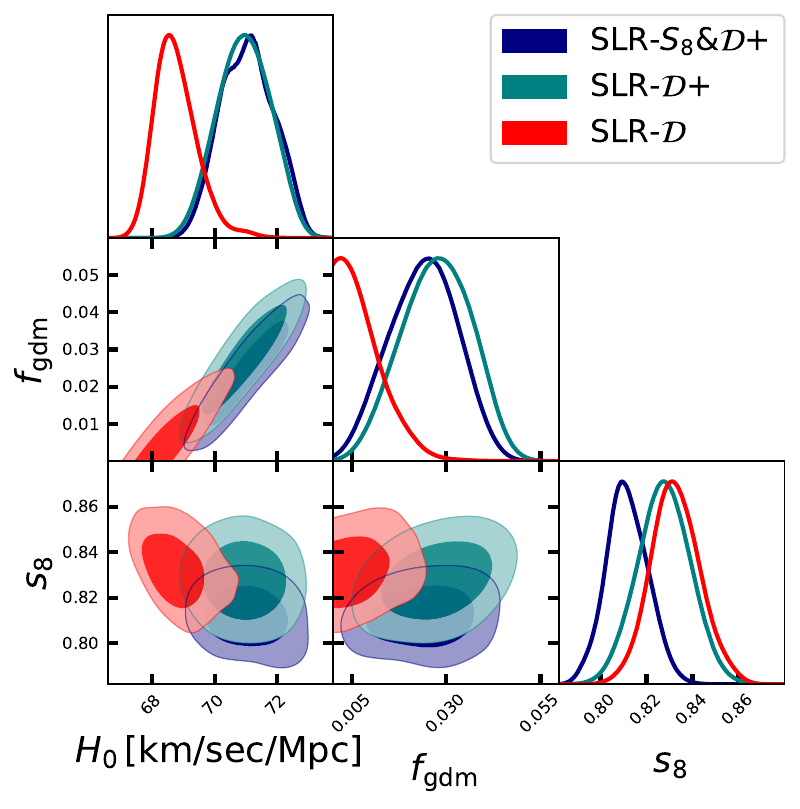}
\caption{\label{fig:s8-triangle} A comparison of the SLR posterior with the $\mathcal{D}$, $\mathcal{D}+$ and $S_8~\&~\mathcal{D}+$ data collections. The shaded regions indicate the 68\% and 95\% credible regions.}
\end{figure}

In another line of inquiry, we see intriguing implications when we look at the impact of $S_8$. Unlike many models proposed in the literature to ease the Hubble tension, there is room within the HDGDM model to increase $ H_0 $ without increasing $ S_8 $. Fig. \ref{fig:s8-triangle} support this idea. In the plot, we post process\footnote{We use importance sampling to add the effect of the $S_8$ prior.} the results of $\mathcal{D}+$ with a prior from DES3 of $S_8=0.776\pm0.017$  \cite{DES:2021wwk}. The figure shows that $S_8$ and $H_0$ are uncorrelated (or even slightly negatively correlated before SH0ES data is added). This is supported by the best-fit values, which suggest that HDGDM can fit all the data in $\mathcal{D}+$ without increasing the $S_8$ tension.

\section{Conclusions}
We have implemented and explored a high-dimensional generalized dark matter (HDGDM) model with a free equation of state $w(a)$. We demonstrated that this model can mimic various cosmological fluids, including models of interest such as typical self-interacting dark radiation species. We applied this formalism to mimic a stepped dark radiation fluid and investigated the constraints of this step-like (SL) model.

We showed that the marginalized posterior intervals are dominated by prior volume effects, leading to potentially rather misleading conclusions about the model viability. In particular, we showed that a more restricted version of the model (SLR) has a much wider range of allowed values for $H_0$ than the unrestricted SL model. Despite adding information through this restriction, the uncertainty on the inferred $H_0$ increases. This serves as a cautionary tail against using the credible intervals of parameters for model selection or even for the search for viable parameter space within a given model.

We have discussed several ways of circumventing these issues, such as using techniques based on minimization or likelihood cutoffs. This also leads us to the conclusion that the $Q_\mathrm{DMAP}$ criterion of \cite{Raveri:2018wln} is an excellent indicator of whether such prior effects are at play. Indeed, another option of estimating whether a model might be useful (or contain a useful subspace of parameters) is to subject it to more challenging data that can direct to a relevant novel region of interest.

By adding a prior on the Hubble constant, for example, we can force the SL and SLR models into a regime where the non-trivial freedom in $w(a)$ can be used more effectively. Indeed, we observe interesting features in the equation of state (a spike at $z \sim 10^{4.5}$ and a dip at $z \sim 10^4$) that motivate future investigation. Furthermore, we show that future data should be capable of either constraining the ability of the HDGDM model to ease the Hubble tension or potentially detecting signatures that differ from \LCDM{}.

We expect the development of fast and robust minimization techniques to aid the community in avoiding misleading conclusions based on marginalized posterior distributions, and stress that the validity of a model should be more robustly assessed. The HDGDM presented in this work serves as a prime example, displaying interesting features which might have otherwise been overlooked and certainly motivate further investigation.

\begin{acknowledgements}
We are grateful to S. Raghunathan, G. Lynch, and K. Prabhu for their assistance with the NFGBS error forecasts. L.~K. and M.~M. acknowledge support from the National Science Foundation of the United States via award \# 2010015 and the United States Department of Energy Office of Science via award DE-SC0009999. N.~S.~acknowledges support from the Maria de Maetzu fellowship grant: CEX2019-000918-M, financiado por MCIN/AEI/10.13039/501100011033.
\end{acknowledgements}
\appendix
\section{Initial Conditions}

To initialize the perturbation variables the asymptotic behavior of the evolution equations is examined. The needed equations are $A1(\mathrm{a-l})$ of \cite{Kopp}.\footnote{The notation of \cite{Ma:1995ey} is used, which differs from \cite{Kopp} and can be converted using $\theta_k = \theta_{\mathrm{MB}}/k^2$ and $\Sigma=\frac{3}{2}\sigma_\mathrm{MB}$} Differing from their approach, set the ansatz of the scale factor to take the form $a=a_i \mathcal{H}_i \tau \sum_{p,q} a_{p,q}\qty(\epsilon_m \tau)^p\qty(\epsilon_g \tau^\beta)^q$, where $\beta=1-3 w_{g,i}$ and $\epsilon_I=\frac{\rho_{I,i}}{\rho_{r,i}}$ for matter and GDM and \mbox{$3 \mathcal{H}_i^2=8 \pi G\rho_{r,i} \cdot a_i^2$} is the conformal expansion rate to zeroth order in matter and GDM. All $i$ subscripts denote the quantities at initial conditions which are assumed to be deep enough in radiation domination to expect $\epsilon_I$ to be small. The treatment assumes that $w_{g}$ is slowly varying at this time period, enough to ignore its evolution while setting initial conditions. The choice of ansatz easily accommodates limiting cases where matter or GDM are turned off.  The metric and materials perturbation variables are expresses in powers of $x=k\tau$. Unfortunately they do not lend themselves to a series in powers of $(\epsilon_m x),(\epsilon_g x^\beta)$ with a finite polynomial in $x$ at zeroth order. Instead express these variable in a series of powers of $x$ and $x^\beta$ which moves the dependence of $\epsilon_g,\epsilon_m$ into the coefficients while admitting the pure radiation solution if desired. The fraction of radiation density that is neutrinos is denoted $ S_\nu = \rho_\nu/\rho_r$.  After using rescaling to set $\eta(x=0)=1$ (used in \class{}) the leading order solutions in small $x$ then become:
\begin{align}
    a&= a_i \mathcal{H}_i \tau\qty(1+\frac{\mathcal{H}_i \tau}{4}\epsilon_m  +\frac{(\mathcal{H}_i\tau)^\beta}{2(1+\beta)}\epsilon_g)~,\\
    \eta&= 1-\frac{5+4S_\nu}{12\qty(15+4S_\nu)}x^2~,\\ h&=\frac{1}{2}x^2~,
\end{align}
for the scale factor and the metric potentials, and
\begin{align}
    \delta_\gamma =\delta_\nu = \frac{4}{3} \delta_b = \frac{4}{3} \delta_c = \frac{-1}{3}x^2~,
\end{align}
as well as
\begin{align}
    \delta_\nu &= \frac{-k(23+4S_\nu)}{36(15+4S_\nu)}x^3~, \\
    \delta_g &=-\frac{\qty(1+w)\qty(1-\frac{3}{4}c_s^2)+12c_v^2\frac{c_s^2-w}{\qty(15+4S_\nu)}}{4+3c_s^2-6w}x^2~,
\end{align}
for the overdensities and
\begin{align}
    \theta_\gamma&=\theta_b=\frac{-k}{36}x^3~,\\
    \theta_g &= \text{\resizebox{18em}{!}{$\frac{-k}{4+3c_s^2-6w}\qty(\frac{c_s^2}{4}+\frac{4c_v^2\qty(2-3\qty(w-c_s^2))}{3\qty(15+4S_\nu)\qty(1+w)})x^3$}}~,
\end{align}
for the velocities and finally
\begin{align}
    \sigma_\nu &= \frac{2 x^2}{3\qty(15+4S_\nu)}~,\\
    \sigma_g &= \frac{8 c_v^2 x^2}{3\qty(15+4S_\nu)\qty(1+w)}~,
\end{align}
for the shear terms. These relations agree with \cite{Kopp} and \cite{Ballesteros:2010} when $w,c_s^2,c_v^2$ are kept to first order and when $c_v^2=0$ respectively. 
\bibliography{HDGDM}
\end{document}